\documentclass[12pt]{iopart}

\usepackage{graphicx}
\usepackage{iopams}
\usepackage{revsymb}

\begin{document}

\title[ S Longhi and G Della Valle, Tamm-Hubbard surface states ...]{Tamm-Hubbard surface states in the continuum}

\author{S Longhi and G Della Valle}

\address{Dipartimento di Fisica, Politecnico di Milano and Istituto di Fotonica e Nanotecnologie del Consiglio Nazionale delle Ricerche, Piazza L. da Vinci
32, I-20133 Milano, Italy}
\ead{longhi@fisi.polimi.it}

\begin{abstract}
\noindent
In the framework of the Bose-Hubbard model, we show that  two-particle surface bound states embedded in the continuum (BIC) can be sustained at the edge of a semi-infinite one-dimensional tight-binding lattice for any infinitesimally-small impurity potential $V$ at the lattice boundary. Such thresholdless surface states, that can be referred to as Tamm-Hubbard BIC states, exist provided that the impurity potential $V$ is attractive (repulsive) and the particle-particle Hubbard interaction $U$ is repulsive (attractive), i.e. for $UV<0$. 
\end{abstract}


\maketitle

\section{Introduction}

Surface waves localized at an
interface between two different media are ubiquitous in several fields of physics  \cite{Davi}. In condensed-matter physics, electronic surface waves 
at the edge of a truncated crystal have been commonly explained as
the manifestation of either Tamm \cite{Tamm} or Shockley \cite{S} localization mechanisms (see, for instance, \cite{Zack}).  Tamm surface states arise from an asymmetrical surface potential
and their formation requires exceeding
a threshold perturbation of the surface potential. On the
other hand,  Shockley surface states result from the crossover of adjacent bands, and can exist
without a surface perturbation. The direct observation of surface states in solids remained elusive
for decades until the advent of semiconductor superlattices \cite{super}.
 In optics, analogues of Tamm and Shockley surface
states have been extensively studied for different types of photonic crystals
and waveguide lattices \cite{uff1,uff2,uff3,uff4,uff5,uff6}, and the role of optical nonlinearities on surface wave localization has been highlighted by several authors (see, e.g., \cite{uff7} and references therein).\\ 
 Surface electronic waves are 
generally regarded as bound states of the single-particle Schr\"{o}dinger equation localized at the edge of a truncated periodic potential with an energy in a gap \cite{boo} (bound states outside the continuum, BOC).  
However, since the original work by  von Neumann and Wigner \cite{Wigner}, it is known that in certain potentials one can find bound (normalizable) states with an energy embedded inside the continuum of scattered states. Bound states in the continuum (BIC) have been generally regarded as fragile states occurring in a few special systems with tailored potential \cite{cazz1a,cazz1b,cazz1c,cazz1d}, generally decaying into resonance states by small perturbations \cite{cazz2} and thus of low physical relevance.  In the simplest case, BIC can arise from destructive quantum interference of the decay channels to the continuum  \cite{cazz3a,cazz3b,cazz3c,cazz3d}, for example by virtue of a simple symmetry constraint \cite{cazz4}. Surface BIC of this kind in a tight-binding lattice model have been suggested, for example, in Ref.\cite{Longhi07}.
In a recent work \cite{Molina12}, Molina and coworkers 
introduced a novel kind of surface Tamm states embedded in the continuum, which     
is structurally robust against perturbations. Topological protection of BIC against the hybridization into the continuum has been also  suggested for a two-dimensional quantum Hall insulator \cite{arx}. The idea of BIC has been recently extended by Zhang and collaborators \cite{Zhang12,Zhang13} to the Hubbard model for two interacting particles hopping on a one-dimensional lattice with an impurity potential. This system represents perhaps the simplest robust  realizations of  a {\it bulk} BIC in an infinitely-extended system sustained by particle correlation.\\
 In this work we show the existence of {\it surface} BIC for the two-particle Bose-Hubbard model on a  semi-infinite one-dimensional tight-binding lattice, that we will refer to as {\it Tamm-Hubbard BIC states}. The main result of our analysis is that  Tamm-Hubbard BIC states, besides to being robust, can be {\it thresholdless}, i.e. they can be found for any infinitesimally small impurity potential $V$ at the edge of the semi-infinite lattice provided that $V$ has opposite sign than the Hubbard energy $U$. This is a distinctive feature as compared to single-particle surface Tamm states and two-particle {\it bulk} BIC, in which a minimum (nonvanishing) threshold value of the impurity potential $V$ is needed to sustain BIC states \cite{note1}. 

\section{Tamm-Hubbard surface states}
Let us consider two (spinless) interacting bosons hopping on a one-dimensional {\it semi-infinite} tight-binding lattice with an impurity at the edge of the lattice.
In the framework of the standard Bose-Hubbard model, the Hamiltonian of the system reads 
 \begin{equation}
 \hat{H}=\sum_{k=0}^{\infty} \left[ - \left( \hat{a}^{\dag}_{k} \hat{a}_{k+1}  + \hat{a}^{\dag}_{k+1} \hat{a}_{k}  \right) +\frac{U}{2} \hat{a}^{\dag 2}_{k}  \hat{a}^{2}_{k} \right]+V  \hat{a}^{\dag }_{0}  \hat{a}_{0}
 \end{equation}
 \begin{figure}
\includegraphics[width=\columnwidth]{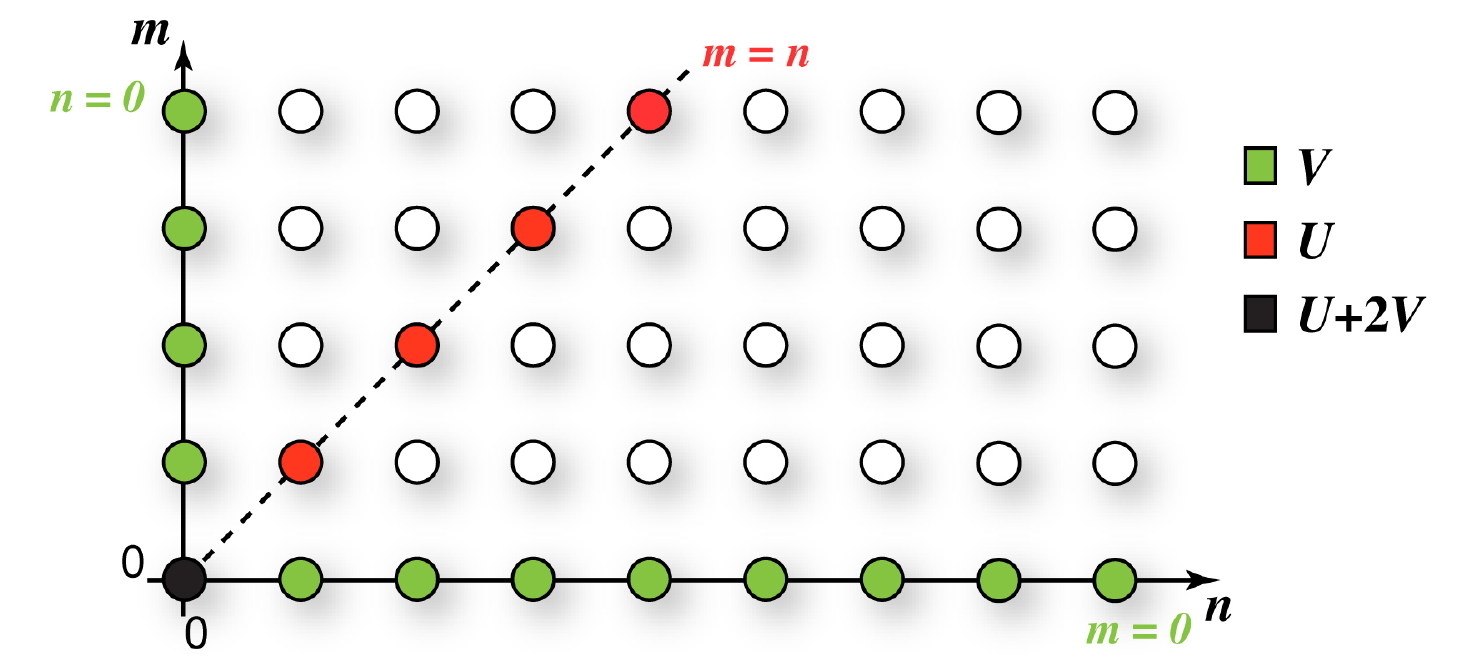}
\caption{(Color online) Schematic of a semi-infinite square lattice that describes in Fock space the dynamics of two interacting bosons hopping on a one-dimensional semi-infinite lattice with an edge impurity potential $V$. The three defect diagonals at $n=0$, $m=0$ and $m=n$ account for the impurity potential $V$ and Hubbard interaction energy $U$.}
\end{figure}
 where $\hat{a}_k$ ($\hat{a}^{\dag}_{k}$) is the annihilation (creation) operator for a boson at lattice site $k$ ($k=0,1,2,...$), $U$ is the on-site particle interaction energy ($U>0$ for a repulsive interaction), and $V$ is the impurity potential energy at the edge lattice site $k=0$ ($V>0$ for a repulsive impurity). In writing Eq.(1), the hopping rate has been set to unity. The Hamiltonian (1) conserves the total number of particles. For the single-particle case, the problem of surface Tamm states is well know: the tight-binding lattice sustains the continuous band $-2<\mathcal{E}<2$ of scattered states, plus one additional eigenvalue, corresponding to a  surface state. The bound state exists provided that $|V|>1$, and its energy is always located outside the band of scattered states. Therefore, as is well-known surface Tamm states for a single particle in a semi-infinite tight-binding lattice are BOC states and show a {\it threshold} of the impurity potential for their existence.\\  
 Let us now discuss the existence of surface states for two interacting particles.  In the two-particle sector of Hilbert space, the state vector $| \psi \rangle$ of the system can be decomposed as 
 $| \psi \rangle = (1/ \sqrt{2})  \sum_{n,m=0}^{\infty} c_{n,m} \hat{a}^{\dag}_n \hat{a}^{\dag}_m | 0 \rangle$, with $c_{n,m}=c_{m,n}$ for bosons.  The amplitudes $c_{n,m}$ define the probabilities to find the two particles at lattice sites $n$ and $m$ \cite{note2}. In the first quantization framework, the spectrum and corresponding eigenstates of $\hat{H}$ are obtained from the Schr\"{o}dinger equation $\hat{H} | \psi \rangle = \mathcal{E} | \psi \rangle$, which yields the following eigenvalue problem for the amplitudes $c_{n,m}$
 \begin{eqnarray}
 \mathcal{E} c_{n,m} & = & - (c_{n+1,m}+c_{n-1,m}+c_{n,m+1}+c_{n,m-1}) \nonumber \\
  & + &  [U \delta_{n,m}+V \delta_{n,0}+V \delta_{m,0}] c_{n,m}
 \end{eqnarray}
with $n,m=0,1,2,3,....$, $c_{n,m}=c_{m,n}$ and with $c_{-1,m}=c_{n,-1} \equiv 0$. Note that Eqs.(2) can be formally viewed as the eigenvalue problem for a {\it single} particle hopping on a {\it two-dimensional} semi-infinite square lattice with an impurity potential along the three semi-infinite lines $n=0$, $m=0$ and $n=m$, as schematically shown in Fig.1. In particular, two-particle {\it surface states} of the Bose-Hubbard Hamiltonian correspond to normalizable two-dimensional surface states in the lattice of Fig.1 which are localized around the corner $n=m=0$. Note that, if $c_{n,m}$ is an eigenstate with energy $\mathcal{E}$ for the Hubbard Hamiltonian with parameters $(U,V)$, then $g_{n,m}=(-1)^{n+m} c_{n,m}$ is an eigenstate with energy $-\mathcal{E}$ for the Hubbard Hamiltonian with parameters $(-U,-V)$. Hence, in the following analysis we can limit to consider  the case $V \geq 0$.  A remarkable property of the eigenvalue problem  (2) is that  exact solutions for the energy spectrum and corresponding eigenstates can be derived analytically \cite{note0}. In the spirit of the Bethe Ansatz, the most general solution to Eqs.(2) can be searched as a superposition of plane waves of the form \cite{Hubb}
\begin{eqnarray}
c_{n,m} & = & A_1 \exp(i k_1n+ik_2m)  +A_2 \exp(-ik_1n+ik_2m) \nonumber \\
& + & A_3 \exp(ik_1n-ik_2m) +A_4 \exp(-ik_1n-ik_2m) \nonumber \\
& + & A_5 \exp(i k_2n+ik_1m)  +A_6 \exp(-ik_2n+ik_1m) \nonumber \\
& + & A_7 \exp(ik_2n-ik_1m) +A_8 \exp(-ik_2n-ik_1m) \;\;\;\;\;\;
\end{eqnarray}
for $n \geq m$, and $c_{n,m}=c_{m,n}$ for $m>n$, 
where $k_1$ and $k_2$ are complex wave numbers and $A_1,A_2,...,A_8$ are eight complex amplitudes. The wave numbers define the energy $\mathcal{E}$ according to the relation
\begin{equation}
\mathcal{E}=-2 \cos k_1-2 \cos k_2
\end{equation}
which is readily obtained after substitution of the Ansatz (3) into Eqs.(2) for $(n,m)$ far from the three defective lines $n=0$, $m=0$ and $n=m$. 
Imposing the validity of Eqs.(2) along these three defective  lines yields a set of 8 homogeneous linear equations for the eight amplitudes $A_l$ ($l=1,2,...,8$), namely $\mathcal{M} \mathbf{v}=0$, where $\mathbf{v}=(A_1,A_2,...,A_8)^T$ and $\mathcal{M}$ is a $8 \times 8$ matrix which is defined in the Appendix. 
It can be shown that $\rm {det} \mathcal{M}=0$, so that the homogeneous linear system $\mathcal{M} \mathbf{v}=0$ admits of a solution for any $k_1$ and $k_2$. The two complex wave numbers $k_1$ and $k_2$ are constrained by the condition that $c_{n,m}$ does not diverge as $n,m \rightarrow \infty$, which ensures the reality of the energy spectrum $\mathcal{E}$. The point spectrum of $\hat{H}$ requires  $\sum_{n,m=0}^{\infty} |c_{n,m}|^2 < \infty$ (and thus necessarily $c_{n,m} \rightarrow 0 $ as $n,m \rightarrow \infty$), whereas the continuous spectrum corresponds to non-normalizable states. In particular, surface two-particle states of the Bose-Hubbard model belong to the point spectrum of $\hat{H}$ and correspond to bound states of the two-dimensional square lattice of Fig.1 localized around the corner $n=m=0$. To determine the spectrum of $\hat{H}$, one should distinguish five different (non-degenerate) cases, which are discussed in detail in the Appendix. The results of the analysis can be summarized as follows.\\
1) The {\it continuous spectrum} comprises two or three bands. The first band, found for real values of $k_1$ and $k_2$, spans the energy interval $(-4,4)$ and corresponds to scattered states where both bosons are delocalized along the semi-infinite lattice. The second band exists for $V>1$ solely and spans  the energy interval $(V+1/V-2,V+1/V+2)$. Physically, this second band describes  one boson trapped by the impurity at the edge $n=0$ of the semi-infinite lattice, whereas the other boson is not trapped by the defect and is delocalized in the lattice. The third band is obtained for complex-conjugate values of $k_1$ and $k_2$, and spans the energy interval $(U, \sqrt{U^2+16})$ (for $U>0$), or $(-\sqrt{U^2+16},U)$ (for $U<0$). This band is the well-known Mott-Hubbard band which describes two-particle bound states (doublons) that undergo correlated tunneling on the lattice (see, for instance, \cite{molecular}).\\
2) The {\it point spectrum}, corresponding to surface states localized near the crystal edge $n=0$, comprises zero, one or two energies, depending on the values of $V$ and $U$. The surface bound states can be either inside (BIC) or outside (BOC) the continuous band $(-4,4)$ of scattered states. The resulting diagram of surface states, either BIC or BOC, is depicted in Fig.2. The equations of the five curves, that determine the boundaries of the existence domains for BIC and BOC, are given in the caption of Fig.2 and are derived in the Appendix.\\ 
 \begin{figure}[t]
\includegraphics[width=\columnwidth]{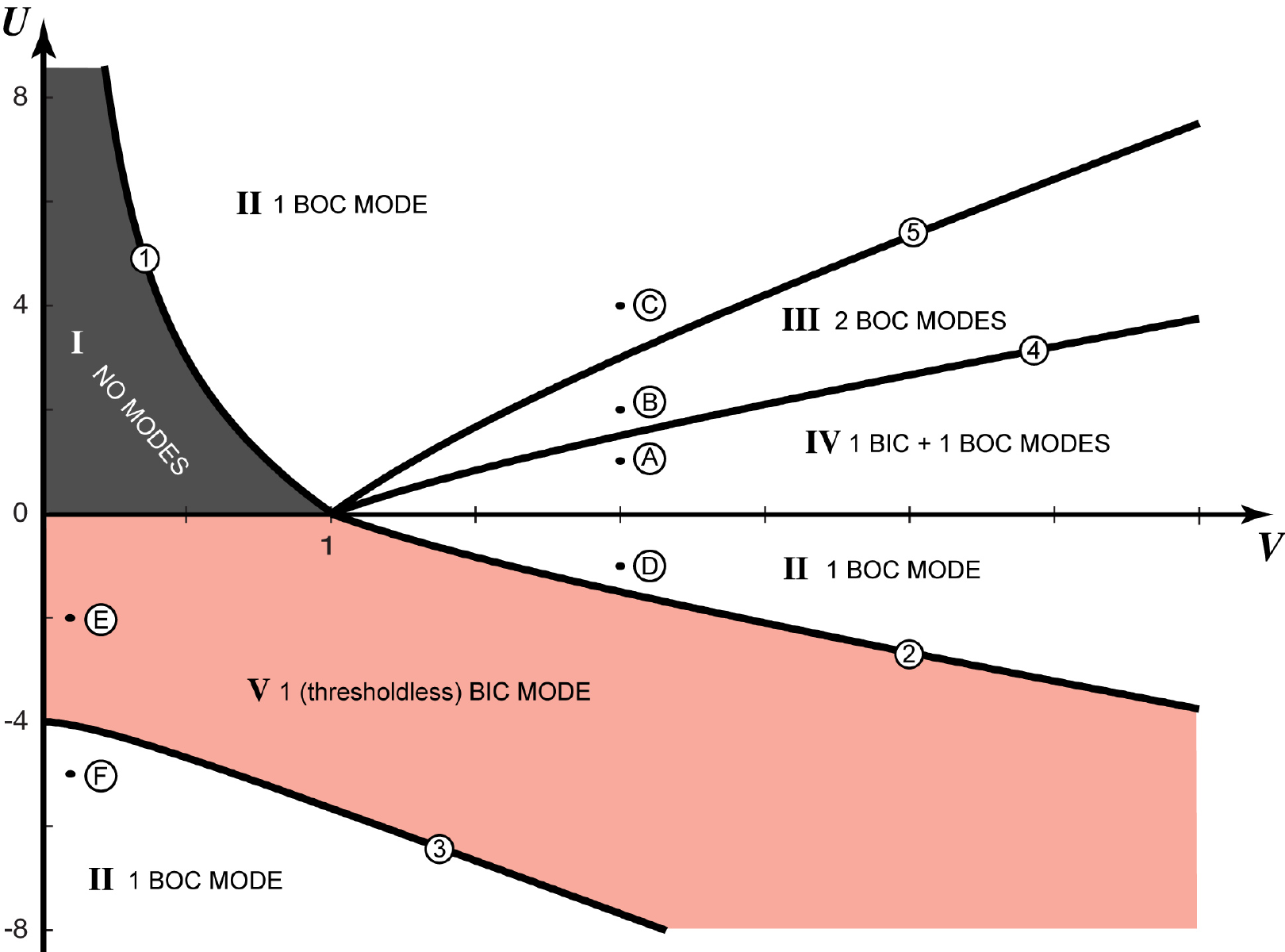}
\caption{(Color online) Existence domains of surface bound states in the $(U,V)$ plane (with $V \geq 0$), either oustide (BOC) or embedded (BIC) into the continuos band $(-4,4)$ of scattered states. In the dark dashed area (domain I) there are not surface states. In the domains II there is one BOC state. In domain III there are two BOC states.  In domain IV there is one BIC and one BOC state. In domain V (light dashed area) there is one BIC. Note that the BIC in the domain V is thresholdless, i.e. it appears for any infinitesimally small value of the potential impurity $V$. The various domains are limited by the five curves shown in the figure. Curve 1: $U=2(1/V-V)$; curve 2: $U=1/V-V$; curve 3: $U=1/V-V- \sqrt{(1/V-V)^2+8/V+8V+16}$; curve 4: $U=V-1/V$; curve 5: $U=2(V-1/V)$. The points A,B,C,D,E and F correspond to the parameter values selected for the numerical simulations shown in Fig.3.}
\end{figure}
 \begin{figure}[b]
\includegraphics[width=\columnwidth]{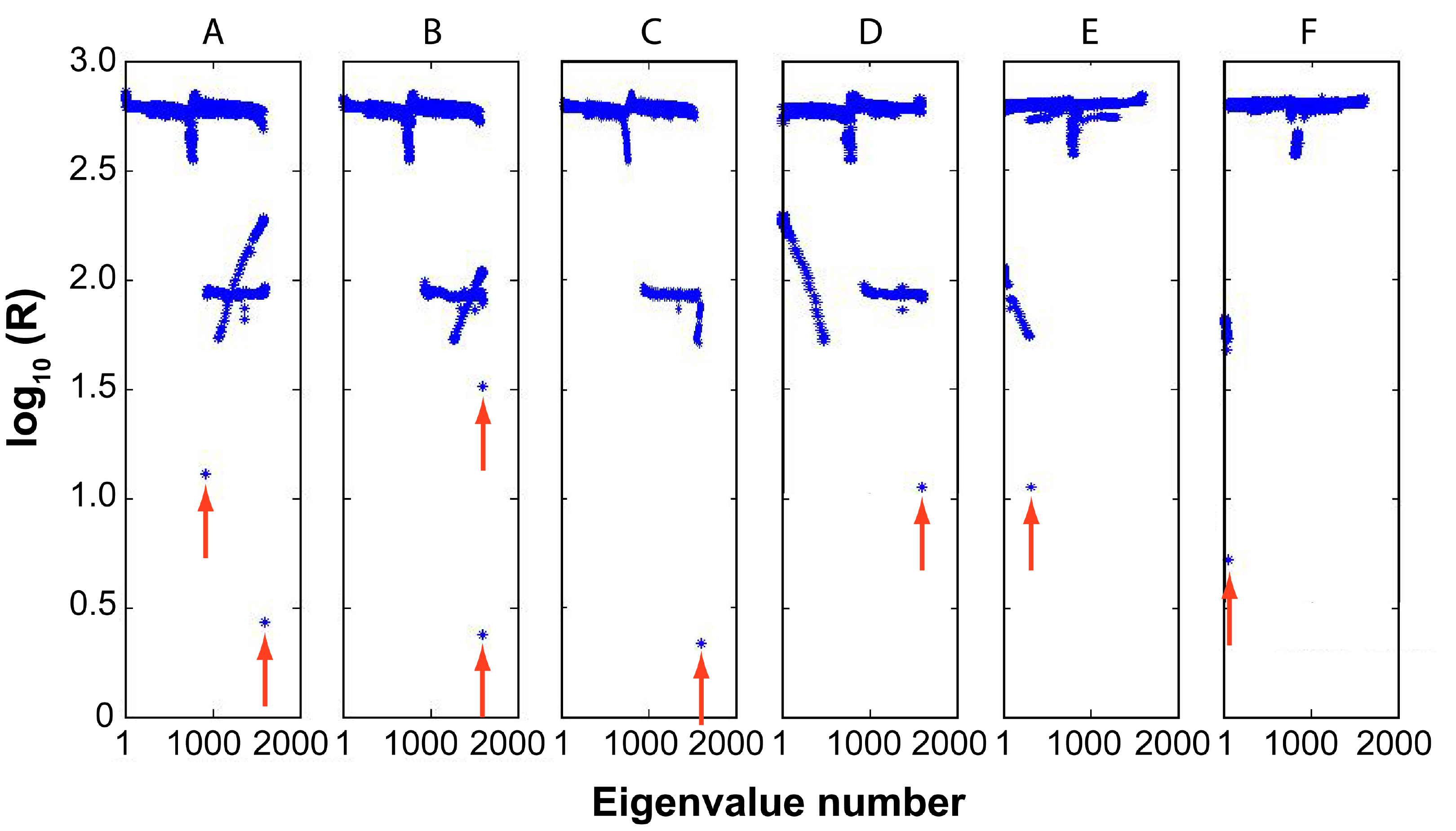}
\caption{(Color online) Numerically-computed logarithm of the participation ratios , $log_{10}(R)$, for the $N \times N=1600$ eigenstates of a Hubbard lattice comprising $N=40$ sites.  Parameter values are $V=2, U=1$ in A, $V=2, U=2$ in B, $V=2, U=4$ in C, $V=2, U=-1$ in D, $V=0.1, U=-2$ in E, and $V=0.1, U=-5$ in F. The arrows in the figures indicate surface bound states.}
\end{figure}
Some important physical results, that follow from an inspection of Fig.2, should be highlighted:\\
(i) For $V=0$ there are not surface bound states, neither embedded nor outside the continuous band of scattered states. Therefore, like for the single particle problem,  in the absence of any potential impurity there are not two-particle surface bound states. Particle interaction solely is not able to sustain a surface state in the absence of a potential impurity at the lattice edge.\\
(ii) For any infinitesimally small value of $V>0$ (repulsive impurity),  there exists a two-particle surface bound state provided that $U<0$ (attractive particle interaction), in spite  Tamm states are not sustained 
 for the single particle. Therefore, contrary to common single-particle Tamm states,  two-particle surface states are {\it thresholdless}. Moreover, if the two particles do not strongly interact, namely for $-4<U<0$, the surface state is a BIC (domain V in Fig.2). Such states can be referred to as {\it Tamm-Hubbard BIC surface states}. Note that Tamm-Hubbard surface states can not be considered as a limiting case of  {\it bulk} BIC of Ref.\cite{Zhang12,Zhang13}. Indeed, the bulk BIC found in \cite{Zhang12,Zhang13} has a finite threshold and is found in the $UV>0$ (rather than $UV<0$) sector of the energy plane \cite{note1}.\\
 (iii) For $V>1$, a  different kind of BIC surface state can be found (domain IV of Fig.2). Note that this surface BIC state has a finite threshold and it always appears in tandem with a BOC surface state.\\
 (iv) There exist two domains in the $(U,V)$ plane where two surface states can be simultaneously sustained. They can be both BOC states (domain III of Fig.2) or one BOC and one BIC state (domain IV in Fig.2).\par
\section{Numerical results}
 We checked the predictions of our analytical results by direct numerical computation of the spectrum and eigenstates of Eqs.(2) in a finite lattice comprising $N=40$ sites, assuming the boundary conditions $c_{n,N}=c_{N,m}=0$ [corresponding to extending the sum in Eq.(1) from $k=0$ to $k=(N-1)$]. The appearance of the surface bound states can be monitored by  computation of the participation ratio, defined by \cite{Molina12} $R=(\sum_{n,m=0, n \neq m}^{N-1} |c_{n,m}|^2)^2 / \sum_{n,m=0, n \neq m}^{N-1} |c_{n,m}|^4$ \cite{note3}. For strongly localized states $R \sim 1$, whereas for extended states $R \sim N^2$.  Lattice truncation at the $n=N$ lattice site does not introduce additional surface states localized near $n=N-1$, the main effect of lattice truncation at the right edge being that of quantizing the wave numbers $k_1$ and $k_2$ for scattered (delocalized) states. Extended numerical simulations spanning the $(U,V)$ plane corroborate the correctness of the bound state diagram of Fig.2. In particular they indicate that  all two-particle bound states of the semi-infinite Bose-Hubbard lattice belong to the states (4) discussed above \cite{note0}.   As an example, in Fig.3 we show the numerically-computed values of ${\rm log}_{10} R$ for six couples of $(U,V)$, corresponding to points A, B, C D, E and F of Fig.2. The arrows in Fig.3 indicate localized surface states. The numerical simulations show that, according to the theoretical predictions, in C, D and F there is one BOC, in B there are two BOC, in A there is one BIC and one BOC, and in E there is one BIC. The latter BIC belongs to the thresholdless BIC discussed above.
  \begin{figure}
\includegraphics[width=12cm]{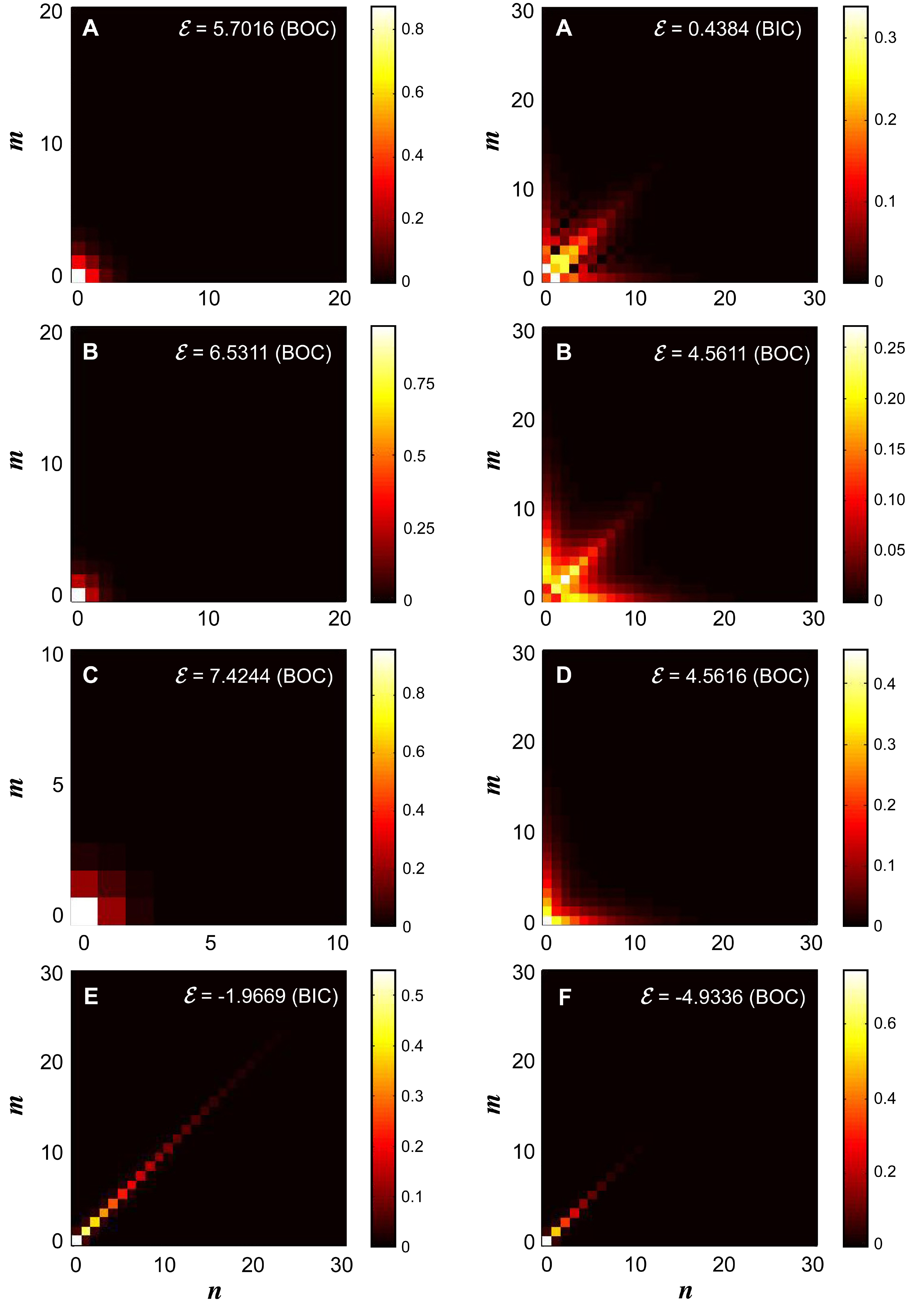}
\caption{(Color online) Numerically-computed distributions of the Fock amplitudes $|c_{n,m}|$ corresponding to surface bound states for parameter values corresponding to points A,B,C,D,E and F of Fig.2. In the figure panels, the energy $\mathcal{E}$ of the surface state is also indicated. BIC and BOC refer to a surface state with energy inside or outside the continuum $(-4,4)$ of scattered states, respectively. Note that in A and B there are two surface bound states.}
\end{figure}
 The distributions $|c_{n,m}|$ for the various surfaces modes, either BOC or BIC, are depicted in Fig.4, where the corresponding energy eigenvalues $\mathcal{E}$ are also indicated. An inspection of Fig.4 clearly shows that, as compared to BOC surface states,  the excitation of the BIC surface modes in Fock space is mostly localized along the diagonal $n=m$. This feature is generally absent for BOC states (compare, for instance, the cases D and E in Fig.4) Physically this means that BOC modes generally correspond to ordinary surface (Tamm) states of two uncorrelated particles, whereas BIC states correspond to a two-particle bound state (doublon), which is localized near the boundary of the lattice owing to the impurity $V$. As $V \rightarrow 0$, the localization length of the two-particle BIC state diverges, and one retrieves the usual delocalized two-particle bound state belonging to the Mott-Hubbard  band embedded into the wider band of uncorrelated (unbounded) particle states.

\section{Conclusion}
In summary, we have predicted a novel type of surface bound states embedded in the continuum  for the two-particle one-dimensional Bose-Hubbard model.  Such states are localized at the edge of a semi-infinite tight-binding lattice and are {\it thresholdess}, i.e. they appear for any infinitesimally-small impurity potential $V$ in the $UV<0$ domain of the energy plane.  Our study  provides what we believe to be the first example of a many-body BIC surface state. We envisage that the present results could be of relevance to different physical fields, ranging from ultacold atoms, quantum dot arrays and photonic waveguide lattices where  the physics of few-particle Hubbard models can be simulated in a controllable way. In particular, two-particle surface BIC states predicted in our work could be observed as surface (corner) states in two-dimensional square lattices of evanescently-coupled optical waveguides with controlled defects \cite{NC}. It would be also interesting to extend our results to the three-particle or many-particle cases. For example, in the two particle case our results show that the Hubbard interaction solely is not able to support any surface BIC state in a one-dimensional truncated lattice, and an impurity potential (even though infinitesimally small) is needed. Would this result break for the many-particle case? Also, our analysis could be extended to investigate many-particle BIC surface states in higher-dimensional lattices or the role of particle correlation in topologically protected bound states in the continuum \cite{arx}.

\appendix
\section{}

In this Appendix we provide a detailed calculation of the spectrum and corresponding eigenstate of Eqs.(2) given in the text.\\
Let us first notice that the Ansatz  given by Eq.(3) satisfies Eqs.(2) far from the three defective  lines $n=0$, $m=0$ and $n=m$ of the lattice of Fig.1, with the energy $\mathcal{E}$ given by Eq.(4).  Imposing the validity of Ansatz (3) at lattice sites $(n,0)$ [or, similarly,  $(0,m)$] and $(n,n)$, i.e. at the defective lines of the square lattice of Fig.1, yields a set of 8 homogeneous linear equations for the eight amplitudes $A_l$ ($l=1,2,...,8$), namely $\mathcal{M} \mathbf{v}=0$, where $\mathbf{v}=(A_1,A_2,...,A_8)^T$ and $\mathcal{M}$ is a $8 \times 8$ matrix, the (nonvanishing) elements of which being given by 
\begin{eqnarray}
\mathcal{M}_{11} & = & \mathcal{M}_{48}=\exp(ik_1)+\exp(-ik_2)-(U-\mathcal{E})/2 \;\;\;\;\;  \nonumber  \\
\mathcal{M}_{15}& = & \mathcal{M}_{44}=\exp(ik_2)+\exp(-ik_1)-(U-\mathcal{E})/2 \;\;\;\;\;  \nonumber \\
\mathcal{M}_{22}& = & \mathcal{M}_{36}=\exp(-ik_1)+\exp(-ik_2)-(U-\mathcal{E})/2 \;\;\;\;\; \nonumber  \\
\mathcal{M}_{27} & = & \mathcal{M}_{33}=\exp(ik_1)+\exp(ik_2)-(U-\mathcal{E})/2 \;\;\;\;\;  \nonumber \\
\mathcal{M}_{51} & = & \mathcal{M}_{62}=V+\exp(-ik_2) \\
\mathcal{M}_{53} & = & \mathcal{M}_{64}=V+\exp(ik_2) \nonumber \\
\mathcal{M}_{75} & = & \mathcal{M}_{86}=V+\exp(-ik_1) \nonumber\\
\mathcal{M}_{77} & = & \mathcal{M}_{88}=V+\exp(ik_1) \nonumber. 
\end{eqnarray}
It can be shown by direct computation that $\rm {det} \mathcal{M}=0$, so that the homogeneous linear system $\mathcal{M} \mathbf{v}=0$ admits of a solution for any complex values of  $k_1$ and $k_2$, defined apart from an unessential multiplication constant. The two complex wave numbers $k_1$ and $k_2$ are constrained by the condition that $c_{n,m}$ does not diverge as $n,m \rightarrow \infty$, which ensures the reality of the energy spectrum $\mathcal{E}$. The point spectrum of $\hat{H}$ requires  $\sum_{n,m=0}^{\infty} |c_{n,m}|^2 < \infty$ (and thus necessarily $c_{n,m} \rightarrow 0 $ as $n,m \rightarrow \infty$), whereas the continuous spectrum corresponds to non-normalizable states. In particular, surface two-particle states of the Bose-Hubbard model belong to the point spectrum of $\hat{H}$ and correspond to bound states of the two-dimensional square lattice of Fig.1(b) localized around the corner $n=m=0$. To determine the spectrum of $\hat{H}$, it turns out that one should distinguish five different (non-degenerate) cases. The first three cases I, II and III determine the continuous spectrum of $\hat{H}$, whereas the last two cases IV and V determine the point spectrum of $\hat{H}$.\\ 
I. {\it $k_1$ and $k_2$ are real-valued}. In this case the energy $\mathcal{E}$, according to Eq.(4), spans the band $(-4,4)$. Correspondingly, the eigenstates are scattered states where both bosons are delocalized along the semi-infinite lattice.\\
II. {\it $ k_1$ is real valued and $k_2$ is imaginary, with ${\rm Im} (k_2)>0$}.  In this case one should take $A_3=A_4=A_6=A_8=0$ to avoid diverging terms in Eq.(3). From the matrix equation $\mathcal{M} \mathbf{v}=0$ it then follows that $V= -\exp(-ik_2)$, so that an acceptable solution, belonging to the continuous spectrum of $\hat{H}$, is found provided that $|V|>1$. According to Eq.(4), the energy band of such states is described by the dispersion relation $\mathcal{E}(k_1)=-2 \cos (k_1)+V+1/V$, which spans the energy interval $(V+1/V-2,V+1/V+2)$. Physically, such states correspond to one boson trapped by the impurity at the edge $n=0$ of the semi-infinite lattice, whereas the other boson is not trapped by the defect and delocalized along the semi-infinite lattice.\\
III. {\it $k_1$ and $k_2$ complex conjugates.} Let us assume $k_1=q+i \rho$ and  $k_2=q-i \rho$, with $\rho>0$ for the sake of definiteness. In this case, to avoid the appearance of diverging terms in Eq.(3) one should assume $A_2=A_4=A_5=A_7=0$. The matrix equation $\mathcal{M} \mathbf{v}=0$ then yields $U=-4 \cos q \sinh \rho$. The corresponding eigenstates are localized at around $n=m$, i.e. $c_{n,m} \rightarrow 0$ as $|n-m| \rightarrow \infty$, however they are delocalized along the diagonal $n=m$ since $c_{n,n}$ does not vanish as $n \rightarrow \infty$. As in previous cases I and II, such states belong to the continuous spectrum of $\hat{H}$ and define a third band with the dispersion relation $\mathcal{E}(q)={\rm sign}(U) \sqrt{U^2+16 \cos^2 q}$. 
This is the well-known  Mott-Hubbard band of particle bound states.
From a physical viewpoint, the Mott-Hubbard band describes molecular bound states (doublons) of the two-particle Hubbard model, in which the two particles form a bound state and hop together along the lattice with an effective hopping rate defined by the bandwidth of the Mott-Hubbard band.\\
IV. {\it $k_1$ and $k_2$ complex valued, with ${\rm Im}(k_1)>0$ and ${\rm Im}(k_1+k_2)>0$}.  In this case, one can assume $A_l=0$ for $l \neq 1$, i.e. $c_{n,m} \sim \exp(ik_1n+ik_2m)$ for $n \geq m$, which corresponds to a surface bound state since $c_{n,m}$ exponentially decays toward zero as $n,m \rightarrow \infty$ . The solvability condition for the matrix equation $\mathcal{M} \mathbf{v}=0$ shows that there exists one acceptable solution, except for the dashed region shown in Fig.2 and delimited by the curves $V=0$, $U=0$ and $U=2(1/V-V)$ (curve 1 in Fig.2). After setting $z_1= \exp(-ik_1)$ and $z_2=\exp(-ik_2)$,  the values of the complex wave numbers for the localized surface state are obtained from the relations $z_2=-V$ and
\begin{equation}
z_1=\frac{1}{2}\left( \frac{1}{V}-V-U \right) \pm \sqrt{ \frac{1}{4}\left( \frac{1}{V}-V-U \right)^2+1}. \label{cazz1}
\end{equation}
The sign in Eq.(\ref{cazz1}) must be chosen such that $|z_1|>max \{1,1/|V| \}$. The surface bound state is a BIC, i.e. its energy $\mathcal{E}=-(z_1+1/z_1+z_2+1/z_2)$ is embedded into the continuos band $(-4,4)$ of scattered states, in the domain shown in Fig.2 and delimited by the curves $U=0$, $V=0$, $U=1/V-V$ (curve 2 in Fig.2), and $U=(1/V-V)-\sqrt{(1/V-V)^2+16+8V+8/V}$  (curve 3 in Fig.2). Such curves are obtained by imposing $\mathcal{E}= \pm 4$ and using the relation $z_2=-V$ and Eq.(S-2).  Note that, for any infinitesimally-small value of $V>0$ and provided that $-4<U<0$, the two-particle surface state is a BIC. Hence this kind of BIC surface state is {\it thresholdless}. \\
V. {\it $k_1$ and $k_2$ complex valued, with ${\rm Im}(k_1)>0$, ${\rm Im}(k_2)<0$ and ${\rm Im}(k_1+k_2)>0$}. In this case to avoid the appearance of secularly growing terms in Eq.(3) one should take $A_l=0$, except for $l=1,3$ and 6. The solvability condition for the matrix equation $\mathcal{M} \mathbf{v}=0$ shows that there exists one surface bound state for $V>1$  in the domains III and IV shown in Fig.2 and delimited by the curves $U=0$ and $U=2(V-1/V)$ (curve 5 in Fig.2). The wave numbers of the surface state are found from the relations $z_1=-V$ and
\begin{equation}
z_2=\frac{1}{2}\left( \frac{1}{V}-V+U \right) \pm \sqrt{ \frac{1}{4}\left( \frac{1}{V}-V+U \right)^2+1}, \label{caz2}
\end{equation}
where $z_1=\exp(-ik_1)$ and $z_2=\exp(-ik_2)$.
 The sign in Eq.(\ref{caz2}) must be chosen such that $1/|z_1|<|z_2|<1$. In particular,  it turns out that the surface state is a BIC in the domain IV delimited by the curves $U=0$ and $U=V-1/V$ (curve 4 in Fg.2).\\
 The domain of existence of surface bound states, either BIC or BOC, is summarized in Fig.2 given in the text.


\begin{thebibliography}{99}
\bibitem{note1}
In a tight-binding lattice, single-particle surface Tamm states exist provided that the impurity energy $V$ at the lattice edge exceeds, in modulus, the particle hopping rate. Two-particle bulk  BIC states predicted Ref.\cite{Zhang12,Zhang13} show a  threshold as well. Indeed, for  a given repulsive (attractive) Hubbard interaction energy $U$, a two-particle bulk BIC exists for a repulsive (attractive) potential impurity $V$ provided that $|V|>|U|$ \cite{Zhang12,Zhang13},.  Another distinctive feature of bulk BIC Hubbard states as compared to BIC Tamm-Hubbard surface states is that the former exist in the $UV>0$ domain, whereas the latter in the  $UV<0$  domain.
\bibitem{note2}
More precisely, $ |c_{n,n}|^2$ is the probability to find the two bosons at the same lattice site $n$, whereas the probability to find one boson at lattice site $n$ and the other one at lattice site $m \neq n$ is given by $ 2|c_{n,m}|^2$.
\bibitem{note0}
For the infinitely-extended Hubbard lattice of Ref.\cite{Zhang12,Zhang13}, the eigenstates can be classified on the basis of their parity (even or odd states) under the inversion $(n,m) \rightarrow (-n,-m)$, and analytical form of the eigenstates can be provided solely for odd-symmetry states, i.e. for states with $c_{-n,-m}=- c_{n,m}$. Conversely, in the semi-infinite lattice considered in our case, where inversion symmetry does not apply, the problem is fully integrable.
\bibitem{note3}
Since bound particle states, belonging to the Mott-Hubbard band, show a strong localization along the diagonal $n=m$, in the computation of the participation ratio $R$ we have excluded from the sums the sites with $n=m$.  In this way localized surface states at the corner $n=m=0$ of the lattice of Fig.1 can be at best distinguished from particle bound states, which are localized on the diagonal $n=m$.
\end{thebibliography}
\end{document}